\documentclass{PoS}

\def\slash#1{\mkern-1.5mu\raise0.4pt\hbox{$\not$}\mkern1.2mu #1\mkern 0.7mu}

\title{Minimally doubled fermions and their renormalization}

\ShortTitle{Minimally doubled fermions}

\author{\speaker{Stefano Capitani}\\
        Institut f\"ur Kernphysik, Becher Weg 45,
        University of Mainz, D-55099 Mainz, Germany\\
        E-mail: \email{capitan@kph.uni-mainz.de}}

\author{Michael Creutz\\
        Physics Department, Brookhaven National Laboratory, Upton,
        NY~11973, USA\\
        E-mail: \email{mike@latticeguy.net}}

\author{Johannes Weber\thanks{Presenter of the poster.}\\
        Graduate School of Pure and Applied Physics, Tsukuba
        University, Tsukuba, Ibaraki, Japan\\
        E-mail: \email{weberj@kph.uni-mainz.de}}

\author{Hartmut Wittig\\
        Institut f\"ur Kernphysik, Becher Weg 45,
        University of Mainz, D-55099 Mainz, Germany\\
        E-mail: \email{wittig@kph.uni-mainz.de}}

\abstract{Minimally doubled fermions have been proposed as a strictly local
discretization of the QCD quark action, which also preserves chiral
symmetry at finite cut-off. We study the renormalization and mixing
properties of two particular realizations of minimally doubled fermions
in lattice perturbation theory at one loop, and we construct conserved 
axial currents which have a simple form involving only nearest-neighbours 
sites. We also introduce a notation which allows a unified description 
of the renormalization properties of both actions.}

\FullConference{The XXVIII International Symposium on Lattice Field Theory, 
                Lattice2010\\
		June 14-19, 2010\\
		Villasimius, Italy}

\begin{document}

\section{Introduction}

We report on our analytic studies of the renormalization properties of 
Bori\c{c}i-Creutz \cite{mind:Creutz07,mind:Borici07,mind:Creutz08,mind:Borici08}
and Karsten-Wilczek \cite{mind:Karsten81,mind:Wilczek87} fermions (see 
\cite{mind:Capitani09,mind:Capitani_lat09,Capitani:2010nn}, and references
therein), two particular realizations of minimally doubled 
fermions.~\footnote{For recent developments, see also \cite{Creutz:2010cz}.}
These actions preserve an exact chiral symmetry for a degenerate doublet 
of quarks, and at the same time they remain strictly local, so that they are 
much cheaper to simulate than Ginsparg-Wilson fermions. They could then 
become a cost-effective realization of chiral symmetry at nonzero lattice 
spacing. This $U(1) \otimes U(1)$ chiral symmetry, which is of the same form
as in the continuum, protects the quark mass from additive renormalization. 
As we have also verified at one loop, the renormalization of the quark 
mass has the same form as, say, overlap or staggered fermions.

It is noteworthy that using minimally doubled fermions one can construct a 
conserved axial current which has a simple expression, involving only 
nearest-neighbour sites (see Section \ref{sec:conscurr}). These actions 
are then among the very few lattice discretizations which provide
a simple (ultralocal) expression for a conserved axial current.

It is natural to compare these realizations of minimally doubled fermions
with staggered fermions, which preserve the same $U(1) \otimes U(1)$ 
chiral symmetry and are also ultralocal and comparably cheap. 
The advantage of Bori\c{c}i-Creutz and Karsten-Wilczek fermions is that
they contain 2 flavours instead of 4, and thus they do not require any 
uncontrolled extrapolation to 2 physical light flavours
\cite{Creutz:2009kx,Creutz:2009zq}. Moreover, the 
construction of fermionic operators is much easier than for staggered fermions, 
where there is also a complicated intertwining of spin and flavour.
Minimally doubled actions look then ideal for $N_f=2$ 
simulations.~\footnote{They remain rather convenient also for $N_f=2+1$ and 
$N_f=2+1+1$ simulations. The second doublet of minimally doubled quarks 
will contain chirality-breaking terms in order to give different masses
to the $s$ and $c$ quarks, however this is not so important for these larger 
masses.}

\section{Actions}
\label{sec:actions}

The free Dirac operator of Bori\c{c}i-Creutz fermions 
is given in momentum space by
\begin{equation}
D(p) = i \, \sum_\mu (\gamma_\mu \sin p_\mu + \gamma'_\mu \cos p_\mu)
- 2i\Gamma +m_0 ,
\label{eq:creutz-action}
\end{equation}
where
\begin{equation}
\Gamma = \frac{1}{2} \, (\gamma_1 + \gamma_2 + \gamma_3 + \gamma_4)
\qquad (\Gamma^2=1)
\end{equation}
and
\begin{equation}
\gamma'_\mu = \Gamma \gamma_\mu \Gamma = \Gamma - \gamma_\mu .
\end{equation}
$D(p)$ vanishes at $p_1=(0,0,0,0)$ and $p_2=(\pi/2,\pi/2,\pi/2,\pi/2)$, and 
can also be seen as a linear combination of two physically equivalent 
naive fermions (one of them translated in momentum space).

The free Karsten-Wilczek Dirac operator is given in momentum space by
\begin{equation}
D(p) = i \sum_{\mu=1}^4 \gamma_\mu \sin p_\mu 
     + i \gamma_4 \sum_{k=1}^3 (1-\cos p_k) ,
\label{eq:wilczek-action}
\end{equation}
and its zeros are instead at $p_1=(0,0,0,0)$ and $p_2=(0,0,0,\pi)$.

The two zeros of these actions, corresponding to the physical flavours,
select a special direction in euclidean spacetime, identified by the line 
that connects them. It is easy to see that in the Bori\c{c}i-Creutz case 
the matrix $\Gamma$ selects as a special direction the major hypercube 
diagonal, while in the Karsten-Wilczek case is the temporal direction 
which becomes the special one. 

As a consequence, hyper-cubic symmetry is broken, and these actions are 
symmetric only under the subgroup of the hyper-cubic group which preserves 
(up to a sign) the respective special direction.
This opens the way to mixings of a new kind under renormalization. 
One of the main aims of our work is the investigation of the mixing patterns 
that appear in radiative corrections. We have elucidated the one-loop 
structure of these theories, and one of our main results is that everything 
is consistent at the one loop level, and the new mixings are very few.

We also remark that, although the distance between the two zeros is the same 
($p_2^2-p_1^2=\pi^2$), these two realizations of minimally doubled fermions 
are not equivalent.

\section{Counterterms}

Each of the two actions (\ref{eq:creutz-action}) and (\ref{eq:wilczek-action})
does not contain all possible operators which are invariant under 
the subgroup of the hyper-cubic group preserving its respective 
special direction. Radiative corrections then generate new contributions 
whose form is not matched by any term in the original bare actions. 
It becomes necessary to introduce counterterms to the bare actions 
in order to obtain a consistent renormalized theory. 
Enforcing the consistency requirement will allow us to uniquely determine 
the coefficients of these counterterms.~\footnote{It is interesting 
in this respect to observe that an action which contains doublers will 
in general select some special direction, and hence require counterterms. 
However, the staggered fermion formulation is very clever, because it 
rearranges the 16 spin-flavour components of the 4 doublers on the corners 
of the unit hypercube. Thanks to this, no special direction arises, and 
thus no extra counterterms are needed for the simulation of staggered fermions.
In the case of naive fermions the 16 doublers are also uniformly distributed 
in the Brillouin zone, and hence there is no special direction 
in this case too.}

One must add to the bare actions all possible counterterms allowed by 
the remnant symmetries. Moreover, counterterms are needed also in 
the pure gauge part of the actions of minimally doubled fermions. 
The reason for this is that, although at the bare level the breaking of 
hyper-cubic symmetry happens only in the fermionic parts of the actions, 
in the renormalized theory it propagates (via the interactions between 
quarks and gluons) also to the pure gauge sector.

We consider the massless case $m_0=0$, and note that chiral symmetry 
strongly restricts the number of possible counterterms.
It turns out that there is only one possible dimension-four fermionic 
counterterm, which for Bori\c{c}i-Creutz fermions is written 
in continuum form as \linebreak
$\overline{\psi} \,\Gamma \sum_\mu D_\mu \psi$. 
A possible discretization for it has a form similar to the hopping term 
in the action:
\begin{equation}
c_4 (g_0) \,\, \frac{1}{2a} \sum_\mu \Big( \overline{\psi} (x) 
\, \Gamma \, U_\mu (x) \, \psi (x + a\widehat{\mu}) 
-\overline{\psi} (x + a\widehat{\mu}) \, \Gamma \,
U_\mu^\dagger (x) \, \psi (x) \Big).
\end{equation}
There is also one counterterm of dimension three,
\begin{equation}
\frac{ic_3 (g_0)}{a}\,\overline{\psi} (x) \, \Gamma \, \psi (x) ,
\end{equation}
which is already present in the bare Bori\c{c}i-Creutz action,
but with a fixed coefficient $-2/a$.
The appearance of this counterterm means that in the general 
renormalized action the coefficient of the dimension-three operator 
must be kept general.

For Karsten-Wilczek fermions we find a similar situation.
The only gauge-invariant fermionic counterterm of dimension four is
\begin{equation}
\overline{\psi}\,\gamma_4 D_4\,\psi , 
\end{equation}
and a suitable discretization of it is
\begin{equation}
d_4 (g_0) \,\, \frac{1}{2a} \Big( \overline{\psi} (x) 
\, \gamma_4 \, U_4 (x) \, \psi (x + a\widehat{4}) 
-\overline{\psi} (x + a\widehat{4}) \, \gamma_4 \,
U_4^\dagger (x) \, \psi (x) \Big) .
\end{equation}
The counterterm of dimension three is for this action
\begin{equation}
\frac{id_3 (g_0)}{a} \,\overline{\psi} (x) \,\gamma_4 \,\psi (x)
\end{equation}
(already present in the bare Karsten-Wilczek action, with a fixed coefficient).

The rules for the counterterm corrections to fermion propagators, needed 
for our one-loop calculations, can be easily derived.
For external lines, they are given in momentum space 
respectively by
\begin{equation}
-ic_4 (g_0) \,\,\Gamma \,\sum_\nu p_\nu , \quad -\frac{ic_3(g_0)}{a}\,\Gamma  
\label{eq:frfctbc}
\end{equation}
for Bori\c{c}i-Creutz fermions, and by
\begin{equation}
-id_4 (g_0) \,\,\gamma_4 \,p_4 , \quad -\frac{id_3(g_0)}{a}\,\gamma_4  
\label{eq:frfctkw}
\end{equation}
for Karsten-Wilczek fermions.

The gluonic counterterms must be of the form $\rm{tr}\, FF$, but with 
nonconventional choices of the indices, reflecting the breaking of the 
hyper-cubic symmetry. It turns out that there is only one purely gluonic 
counterterm, which for the Bori\c{c}i-Creutz action can be written in
continuum form as
\begin{equation}
c_P(g_0) \, \sum_{\lambda\rho\tau} \rm{tr}\, F_{\lambda\rho}(x) \, F_{\rho\tau}(x) .
\end{equation} 
At one loop this counterterm is relevant only for gluon propagators.
Denoting the fixed external indices at their ends with $\mu$ and 
$\nu$, all possible lattice discretizations of this counterterm 
give in momentum space the same Feynman rule:
\begin{equation}
-c_P(g_0) \,\left[ (p_\mu + p_\nu)\,\sum_\lambda p_\lambda -p^2 
- \delta_{\mu\nu}\Big( \sum_\lambda p_\lambda \Big)^2 \right] .
\label{eq:frgctbc}
\end{equation}
Contributions of this kind must be taken into account for a correct 
renormalization of the vacuum polarization (see Section \ref{sec:detgluon}).

In the case of Karsten-Wilczek fermions the counterterm which needs to be 
introduced can be written in continuum form as
\begin{equation}
d_P(g_0) \, \sum_{\rho\lambda} \rm{tr}\, F_{\rho\lambda}(x) \, F_{\rho\lambda}(x) 
\, \delta_{\rho 4} .
\end{equation}
The Feynman rule for the insertion of this counterterm in external gluon
propagators reads
\begin{equation}
- d_P(g_0) \, \left[ p_\mu p_\nu \,(\delta_{\mu 4 } + \delta_{\nu 4 }) 
-\delta_{\mu\nu} \left( p^2\,\delta_{\mu 4 } \delta_{\nu 4 } +p_4^2 \right) 
\right] .
\label{eq:frgctkw}
\end{equation}

In perturbation theory the coefficients of all counterterms 
are functions of the coupling which start at order $g_0^2$.
We will determine (at one loop) the coefficients of all fermionic and 
gluonic counterterms by requiring that the renormalized self-energy and 
vacuum polarization, respectively, assume their standard form
(see Sections \ref{sec:detfermion} and \ref{sec:detgluon}).

Counterterm interaction vertices are generated as well.
However, these vertex insertions are at least of order $g_0^3$, and thus 
they cannot contribute to the one-loop amplitudes that we study here.
We also want to emphasize that counterterms not only provide 
additional Feynman rules for the calculation of loop amplitudes.
They can also modify Ward identities and hence, in particular, contribute 
additional terms to the conserved currents (see Section \ref{sec:conscurr}).

\section{Determination of the fermionic counterterms}
\label{sec:detfermion}

Leaving aside for one moment the counterterms, the quark self-energy 
of a Bori\c{c}i-Creutz fermion is given at one loop by 
\begin{equation}
\Sigma (p,m_0) = i\slash{p}\,\Sigma_1(p) +m_0\,\Sigma_2(p) 
+ c_1 (g_0)\cdot i\, \Gamma \sum_\mu p_\mu
+ c_2 (g_0)\cdot i\, \frac{\Gamma}{a},
\label{eq:totalselfbc}
\end{equation}
where~\footnote{For our calculations we have developed programs written 
in the algebraic computer language {\em FORM} 
\cite{Vermaseren:2000nd,Vermaseren:2008kw}.}
\begin{eqnarray}
\Sigma_1(p) &=& \frac{g_0^2}{16\pi^2} \,C_F \,\Bigg[ \log a^2p^2
  +6.80663 +(1-\alpha) \Big(-\log a^2p^2 + 4.792010 \Big) \Bigg] ,
\label{eq:Sigma1self} \\ 
\Sigma_2(p) &=& \frac{g_0^2}{16\pi^2} \,C_F \,\Bigg[ 4\,\log
  a^2p^2 -29.48729 +(1-\alpha) \Big(-\log a^2p^2 +5.792010 \Big) \Bigg] ,
\label{eq:Sigma2self} \\ 
c_1 (g_0)\, &=&  1.52766 \cdot\frac{g_0^2}{16\pi^2} \,C_F ,
 \label{eq:c1self} \\
c_2 (g_0)\, &=& 29.54170 \cdot\frac{g_0^2}{16\pi^2} \,C_F ,
\end{eqnarray}
with $C_F=(N_c^2-1)/2N_c$, and $\alpha$ denotes the gauge parameter
in a general covariant gauge.
The full inverse propagator at one loop can be written 
(without counterterms) as
\begin{equation}
\Sigma^{-1} (p,m_0) = \Big( 1 -\Sigma_1 \Big) \cdot
\Big\{ i\slash{p}
+ m_0 \,\Big( 1 -\Sigma_2 +\Sigma_1 \Big)
-ic_1 \,\Gamma \,\sum_\mu p_\mu
-\frac{ic_2}{a}\,\Gamma \Big\} .
\end{equation}
We can only cast the renormalized propagator in the standard form
\begin{equation}
\Sigma (p,m_0) = \frac{Z_2}{i\slash{p} + Z_m\, m_0} ,
\end{equation}
where the wave-function and quark mass renormalization factors are given by
\begin{equation}
Z_2 = \Big( 1 -\Sigma_1 \Big)^{-1}, \qquad
Z_m = 1 - \Big( \Sigma_2 -\Sigma_1 \Big) ,
\end{equation}
provided that we employ the counterterms to cancel the Lorentz non-invariant 
factors ($c_1$ and $c_2$).

The term proportional to $c_1$ can be eliminated 
by using the dimension-four counterterm, 
$\overline{\psi} \, \Gamma \, \sum_\mu D_\mu \, \psi$,
while the term proportional to $c_2$ can be eliminated using
the dimension-three counterterm, 
$1/a \, \overline{\psi} \, \Gamma \, \psi$.
This amounts to applying the insertions of eqs.~(\ref{eq:frfctbc})
and (\ref{eq:frfctkw}). We thus determine in this way that at one loop,
for Bori\c{c}i-Creutz fermions,
\begin{equation}
    c_3 (g_0) = 29.54170\cdot\frac{g_0^2}{16\pi^2} \,C_F+O(g_0^4), \qquad
    c_4 (g_0) =  1.52766\cdot\frac{g_0^2}{16\pi^2} \,C_F+O(g_0^4) .
\end{equation} 

Things work out very similarly for Karsten-Wilczek fermions. In this case
the inverse propagator at one loop (without counterterms) is
\begin{equation}
\Sigma^{-1} (p,m_0) = \Big( 1 -\Sigma_1 \Big) \cdot
\Big( i\slash{p} + m_0 \,\Big( 1 -\Sigma_2 +\Sigma_1 \Big)
- id_1 \,\gamma_4 p_4 - \frac{id_2}{a} \,\gamma_4 \Big),
\label{eq:totalselfkw}
\end{equation}
where
\begin{eqnarray}
\Sigma_1(p) &=& \frac{g_0^2}{16\pi^2} \,C_F \,\Bigg[ \log a^2p^2
  +9.24089 +(1-\alpha) \Big(-\log a^2p^2 + 4.792010 \Big) \Bigg] ,
\label{eq:Sigma1self2} \\ 
\Sigma_2(p) &=& \frac{g_0^2}{16\pi^2} \,C_F \,\Bigg[ 4\,\log
  a^2p^2 -24.36875 +(1-\alpha) \Big(-\log a^2p^2 +5.792010 \Big) \Bigg] ,
\label{eq:Sigma2self2} \\ 
d_1 (g_0)\, &=&  -0.12554 \cdot\frac{g_0^2}{16\pi^2} \,C_F ,
 \label{eq:c1self2} \\
d_2 (g_0)\, &=& -29.53230 \cdot\frac{g_0^2}{16\pi^2} \,C_F .
\end{eqnarray}
By using the appropriate counterterms 
$\overline{\psi} \, \gamma_4 \, D_4 \, \psi$ and
$1/a \, \overline{\psi} \, \gamma_4 \, \psi$
the renormalized propagator can be written in the standard form.
Then, at one loop we obtain
\begin{equation}
    d_3 (g_0) = -29.53230\cdot\frac{g_0^2}{16\pi^2} \,C_F+O(g_0^4), \qquad
    d_4 (g_0) =  -0.12554\cdot\frac{g_0^2}{16\pi^2} \,C_F+O(g_0^4) .
\end{equation}

One may expect that the above subtraction procedure can be carried out
systematically at every order of perturbation theory. After the
subtractions via the appropriate counterterms are properly taken into
account, the extra terms appearing in the self-energy disappear.

\section{Determination of the gluonic counterterms}
\label{sec:detgluon}

Leaving aside for one moment the counterterms, the contribution of the 
fermionic loops to the one-loop vacuum polarization of Bori\c{c}i-Creutz 
fermions comes out from our calculations as
\begin{eqnarray}
\Pi^{(f)}_{\mu\nu} (p) & = & \Bigg( p_\mu p_\nu-\delta_{\mu\nu}p^2 \Bigg) 
\Bigg[\frac{g_0^2}{16\pi^2} C_2 \Bigg( -\frac{8}{3} \log p^2a^2 + 23.6793
\Bigg) \Bigg] \\
&& - \Bigg( (p_\mu + p_\nu)\,\sum_\lambda p_\lambda - p^2
- \delta_{\mu\nu}\Big( \sum_\lambda p_\lambda \Big)^2 \Bigg) \,
\frac{g_0^2}{16\pi^2} \,C_2 \cdot 0.9094 , \nonumber
\end{eqnarray}
where $\rm{Tr} \,(t^at^b) = C_2 \,\delta^{ab}$.
For Karsten-Wilczek fermions the corresponding result is
\begin{eqnarray}
\Pi^{(f)}_{\mu\nu} (p) & = & \Bigg( p_\mu p_\nu-\delta_{\mu\nu}p^2 \Bigg) 
\Bigg[\frac{g_0^2}{16\pi^2} C_2 \Bigg( -\frac{8}{3} \log p^2a^2 + 19.99468
\Bigg) \Bigg] \\
&& - \Bigg( p_\mu p_\nu \,(\delta_{\mu 4 } + \delta_{\nu 4 }) 
-\delta_{\mu\nu} \left( p^2\,\delta_{\mu 4 } \delta_{\nu 4 } +p_4^2 \right) \Bigg)\,
\frac{g_0^2}{16\pi^2} \,C_2 \cdot 12.69766 . \nonumber
\end{eqnarray}
We notice the appearance of non-standard terms, compared with e.g. Wilson 
fermions. These new terms break hyper-cubic symmetry. It is remarkable that 
they still satisfy the Ward identity $p^\mu \Pi^{(f)}_{\mu\nu} (p)=0$.

At this stage we can employ the gluonic counterterms, which correspond to
the insertions in the gluon propagator according to eqs.~(\ref{eq:frgctbc}) 
and (\ref{eq:frgctkw}), to cancel the hyper-cubic-breaking terms 
in the vacuum polarization. 
The coefficients of these counterterms are hence determined as
\begin{equation}
    c_P (g_0) =  -0.9094  \cdot\frac{g_0^2}{16\pi^2} \,C_2 +O(g_0^4), \qquad
    d_P (g_0) = -12.69766 \cdot\frac{g_0^2}{16\pi^2} \,C_2 +O(g_0^4) .
\end{equation}

It is also very important to remark that no power-divergences
($1/a^2$ or $1/a$) show up in our results for the vacuum polarization.

\section{Conserved currents}
\label{sec:conscurr}

We have also calculated the renormalization of the local Dirac bilinears.
We have found that no mixings occur for the scalar and pseudoscalar densities 
and the tensor current. For the vector and axial currents instead a mixing 
can be seen, which is a consequence of the breaking of hyper-cubic invariance,
and their renormalization factors $Z_V$ and $Z_A$ are thus are not equal to one
(for their numerical values see Section \ref{sec:notation}). 
These local currents are indeed not conserved. Using chiral Ward identities
we have then derived the expressions of the conserved currents, which 
are protected from renormalization.

As we have previously remarked, the counterterms influence the expressions 
of the conserved currents. It is easy to see that the counterterm of 
dimension three does not modify the Ward identities, and is irrelevant in
this regard. On the contrary, the dimension-four counterterm 
\begin{equation}
\frac{c_4(g_0)}{4} \sum_\mu \sum_\nu \Big( \overline{\psi} (x) 
\, \gamma_\nu \, U_\mu (x) \, \psi (x + a\widehat{\mu}) 
+\overline{\psi} (x + a\widehat{\mu}) \, \gamma_\nu \,
U_\mu^\dagger (x) \, \psi (x) \Big)
\end{equation}
generates new terms in the Ward identities and hence contributes to 
the conserved currents.
The conserved axial current for Bori\c{c}i-Creutz fermions 
in the renormalized theory turns out to have the expression
\begin{eqnarray}
A_\mu^{\mathrm c} (x) &=& \frac{1}{2} \bigg(
   \overline{\psi} (x) \, (\gamma_\mu+i\,\gamma'_\mu) \gamma_5 \,
   U_\mu (x) \,
   \psi (x+a\widehat{\mu})
  + \overline{\psi} (x+a\widehat{\mu}) \, (\gamma_\mu-i\,\gamma'_\mu)
   \gamma_5
   \, U_\mu^\dagger (x) \, \psi (x) \bigg) \nonumber \\
 && +\frac{c_4 (g_0)}{2} \,
 \bigg( \overline{\psi} (x) \, \Gamma \gamma_5 \, 
   U_\mu (x) \, \psi (x+a\widehat{\mu})  
  + \overline{\psi} (x+a\widehat{\mu}) \, \Gamma \gamma_5 \, 
   U_\mu^\dagger (x) \, \psi (x) \bigg) .  
\label{eq:noether-axial}
\end{eqnarray}
For Karsten-Wilczek fermions, application of the chiral Ward identities 
gives for the conserved axial current
\begin{eqnarray}
A_\mu^{\mathrm c} (x) & = & \frac{1}{2} \bigg(
   \overline{\psi} (x) \, (\gamma_\mu -i\gamma_4 \, (1-\delta_{\mu 4}) ) \,
   \gamma_5 \, U_\mu (x) \,
   \psi (x+a\widehat{\mu}) \nonumber \\
 && \qquad + \overline{\psi} (x+a\widehat{\mu}) \,
    (\gamma_\mu +i\gamma_4 \, (1-\delta_{\mu 4}) ) \, \gamma_5
   \, U_\mu^\dagger (x) \, \psi (x) \bigg) \\
 && + \frac{d_4 (g_0)}{2} \bigg(
   \overline{\psi} (x) \, \gamma_4 \gamma_5 \, U_4 (x) \,
   \psi (x+a\widehat{4})
  + \overline{\psi} (x+a\widehat{4}) \, \gamma_4 \gamma_5
   \, U_4^\dagger (x) \, \psi (x) \bigg) . \nonumber 
\end{eqnarray}
The conserved vector currents can be obtained by simply dropping 
the $\gamma_5$ matrices from the above expressions.
We remark that the vector current is isospin-singlet, representing the 
conservation of fermion number (as also discussed in \cite{Tiburzi:2010bm}). 
The axial current, however, is a non-singlet because the doubled fermions 
have opposite chirality.  
All these currents have a very simple structure, which involves only 
nearest-neighbour sites.

We have computed the renormalization of these point-split currents,
and verified that is one. As all four cases are very similar, 
we briefly discuss here the conserved vector current for Bori\c{c}i-Creutz 
fermions, for which the sum of the ``standard'' diagrams (vertex, sails and 
operator tadpole, without the counterterm) gives 
\begin{equation}
\frac{g_0^2}{16\pi^2} \,C_F \,\gamma_\mu\,\Bigg[ -\log a^2p^2 -6.80664
+(1-\alpha) \Big(\log a^2p^2 -4.79202 \Big) \Bigg] 
+c_1^{cv} (g_0) \, \Gamma .
\end{equation}
The value of the coefficient of the mixing is
$c_1^{cv} (g_0) \, = -1.52766 \cdot \frac{g_0^2}{16\pi^2} \,C_F + O(g_0^4)$.

When one adds to this result the wave-function renormalization 
(that is, $\Sigma_1(p)$ of the quark self-energy), the term proportional 
to $\gamma_\mu$ is exactly cancelled. The mixing term, proportional to 
$\Gamma$, instead remains, because we have not yet taken into account 
the counterterm.

The part of the conserved vector current due to the counterterm
corresponds to the last line of eq.~(\ref{eq:noether-axial}).
Its 1-loop contribution is quite easy to compute (since $c_4$ is already of 
order $g_0^2$), and is given by $c_4(g_0) \, \Gamma$.
We now note that the value of $c_4$ is already known from the self-energy, 
and numerical inspection shows that $c_4(g_0) = -c_1^{cv}(g_0)$ (within
the precision of our integration routines).
Thus, the $\Gamma$ mixing term is finally cancelled. We emphasize that only 
this particular value of $c_4$, determined from the self-energy, does exactly 
this job.

We have thus obtained that the renormalization constant of these 
point-split currents is one, which confirms that they are conserved currents.
Everything turns out to be consistent at the one loop level.

\section{Numerical simulations}
\label{sec:simulations}

If we use the nearest-neighbour forward covariant derivative 
$\nabla_\mu \psi(x) = 
\frac{1}{a}\,[U_\mu(x)\,\psi\,(x+a\widehat{\mu}) - \psi(x)]$ and 
the corresponding backward one $\nabla^\ast_\mu$, we can express the
(bare) actions in position space in a rather compact form. It then becomes 
apparent that these two realizations of minimally doubled fermions 
bear a close formal resemblance to Wilson fermions:
\begin{eqnarray}
D^f_{\rm{Wilson}}  & = & \frac{1}{2} \, \Bigg\{
\sum_{\mu=1}^4 \gamma_\mu (\nabla_\mu + \nabla^\ast_\mu) \,
-ar \sum_{\mu=1}^4 \nabla^\ast_\mu \nabla_\mu \Bigg\} , \\
D^f_{\rm{BC}} & = & \frac{1}{2} \, \Bigg\{
\sum_{\mu=1}^4 \gamma_\mu (\nabla_\mu + \nabla^\ast_\mu) \,
+ia \sum_{\mu=1}^4 \gamma'_\mu \,\nabla^\ast_\mu \nabla_\mu \Bigg\} , \\
D^f_{\rm{KW}} & = & \frac{1}{2} \, \Bigg\{
\sum_{\mu=1}^4 \gamma_\mu (\nabla_\mu + \nabla^\ast_\mu) \,
-ia \gamma_4 \sum_{k=1}^3 \nabla^\ast_k \nabla_k \Bigg\} .
\end{eqnarray}
All these three formulations contain a dimension-five operator in the bare 
action, and so we expect leading lattice artefacts to be of order~$a$.
However, for minimally doubled fermions these effects could numerically 
be small, if the results of \cite{Cichy:2008gk} are to be believed.

We will not discuss here how to achieve one-loop (or nonperturbative) 
order~$a$ improvement for these theories. The classification of all relevant 
independent operators could turn out to require a lengthy analysis.
Notice that additional dimension-5 operators will occur not only in the quark 
sector (e.g., $\overline{\psi}\,\Gamma \sum_{\mu\nu}D_\mu D_\nu \psi$), 
but also in the pure gauge part 
(e.g., $\sum_{\mu\nu\lambda}F_{\mu\nu}D_\lambda F_{\mu\nu}$).
Indeed, when Lorentz invariance is broken, the statement that only operators
with even dimension can appear in the pure gauge action is no longer true.

We would now like to see what can be learned, from the one-loop calculations 
that we have carried out, regarding the numerical simulations of minimally 
doubled fermions. These simulations will have to employ the complete 
renormalized actions, including the counterterms. 

The renormalized action for Bori\c{c}i-Creutz fermions in position space 
contains three counterterms and reads
\begin{eqnarray}
S^f_{BC} & = & a^4 \sum_{x} \bigg\{ \frac{1}{2a} \sum_{\mu=1}^4 \Big[
    \overline{\psi} (x) \, (\gamma_\mu + c_4(\beta) \, \Gamma 
           + i\gamma'_\mu) \, U_\mu (x) \, \psi (x + a\widehat{\mu}) 
\nonumber \\
&& \qquad -\overline{\psi} (x + a\widehat{\mu}) \, (\gamma_\mu 
      + c_4(\beta) \, \Gamma - i\gamma'_\mu) \,
   U_\mu^\dagger (x) \, \psi (x) \Big] \nonumber \\
&& \qquad  + \overline{\psi}(x) \, \Big(m_0+\widetilde{c}_3(\beta)\,
    \frac{i\,\Gamma}{a} \Big) \, \psi (x)  
\nonumber \\
&& \qquad +\beta \sum_{\mu < \nu} \Bigg( 1 - \frac{1}{N_c} 
    {\mathrm Re} \, \rm{tr}\, P_{\mu\nu} \Bigg)
  + c_P(\beta) \, \sum_{\mu\nu\rho} \rm{tr}\, 
        F^{lat}_{\mu\rho}(x) \, F^{lat}_{\rho\nu}(x) \bigg\} ,
\end{eqnarray}
where $F^{lat}$ is some lattice discretization of the field-strength tensor.
We have here redefined the coefficient of the dimension-3 counterterm, 
using $\widetilde{c}_3(\beta)=-2+c_3(\beta)$ 
(which does not vanish at tree level).~\footnote{We assume that simulations
will be carried out at very small values of $m_0$, so that our analysis of 
the counterterms, which assumes chiral symmetry, is essentially still valid. 
But note also that in our results of 
eqs.~\ref{eq:totalselfbc} and \ref{eq:totalselfkw}, obtained for general 
$m_0$, no new dimension-four terms proportional to this mass appear
(apart from the standard one, $\Sigma_2$). Thus, at one loop we do not need 
further counterterms in additions to the three which we have found. 
This strongly suggests that our analysis of the counterterms remains valid 
even when chiral symmetry is broken.}

The renormalized action for Karsten-Wilczek fermions also 
contains three counterterms and reads
\begin{eqnarray}
\!\!\!\!\!\!\!\!\!\!\!
S^f_{KW} & = & a^4 \sum_{x} \bigg\{ \frac{1}{2a} \sum_{\mu=1}^4 \Big[
    \overline{\psi} (x) \, (\gamma_\mu(1 + d_4(\beta)\,\delta_{\mu 4}) 
  -i\gamma_4 \, (1-\delta_{\mu 4}) ) \, U_\mu (x) \, \psi (x + a\widehat{\mu}) 
\nonumber \\
&& \qquad -\overline{\psi} (x + a\widehat{\mu}) \, 
(\gamma_\mu(1 + d_4(\beta)\,\delta_{\mu 4}) +i\gamma_4 \, (1-\delta_{\mu 4}) ) \, 
  U_\mu^\dagger (x) \, \psi (x) \Big] \nonumber \\
&& \qquad + \overline{\psi}(x) \, \Big(m_0+\widetilde{d}_3(\beta)\,
  \frac{i\,\gamma_4}{a}\Big) \, \psi (x) \nonumber \\
&& \qquad   + \beta \sum_{\mu < \nu} \Bigg( 1 - \frac{1}{N_c} {\mathrm Re} 
   \, \rm{tr} \, P_{\mu\nu} \Bigg)
    \, \Big( 1 + d_P(\beta) \, \delta_{\mu 4} \Big) \bigg\}  
\end{eqnarray}
($\widetilde{d}_3(\beta)=3+d_3(\beta)$ has a non-zero value 
at tree level).

In perturbation theory the coefficients of the counterterms
have the expansions
\begin{eqnarray}
\widetilde{c}_3(g_0) & = -2 + c_3^{(1)} g_0^2 + c_3^{(2)} g_0^4 + \dots; 
\qquad
\widetilde{d}_3(g_0) & = 3 + d_3^{(1)} g_0^2 + d_3^{(2)} g_0^4 + \dots \\
c_4(g_0) & = \phantom{-2 +~} c_4^{(1)} g_0^2 + c_4^{(2)} g_0^4 + \dots; 
\qquad
d_4(g_0) & = \phantom{3 +~} d_4^{(1)} g_0^2 + d_4^{(2)} g_0^4 + \dots \\
c_P(g_0) & = \phantom{-2 +~} c_P^{(1)} g_0^2 + c_P^{(2)} g_0^4 + \dots; 
\qquad
d_P(g_0) & = \phantom{3 +~} d_P^{(1)} g_0^2 + d_P^{(2)} g_0^4 + \dots .
\end{eqnarray}
The same counterterms also appear at the nonperturbative level, 
and need to be taken into account for a consistent simulation of these 
fermions. Their nonperturbative determination is one the most important 
task for the near future. This can be achieved using suitable 
renormalization conditions, and it remains to be seen which ones will 
turn out to be more convenient in practice.

We have previously seen that in perturbation theory the four-dimensional 
fermionic counter\-term is necessary for the proper construction of the 
conserved currents. Its coefficient, as determined from the one-loop 
self-energy, has exactly the right value for which the conserved currents 
remain unrenormalized. This suggests that one possible nonperturbative 
determination of $c_4$ (and $d_4$) can be accomplished by simulating matrix 
elements of the (unrenormalized) conserved current, and imposing 
(by tuning the coefficient) that the electric charge is one.

Another effect of radiative corrections is to move the poles 
of the quark propagator away from their tree-level positions.
It is the task of the dimension-three counterterm, for the 
appropriate value of the coefficient $c_3$ (or $d_3$), 
to bring the two poles back to their original locations.
These shifts of the poles can introduce oscillations in some hadronic 
correlation functions as a function of time separation (similarly to staggered 
fermions). Then one possible way to determine $c_3$ ($d_3$) is to tune it in 
appropriately chosen correlation functions until these oscillations are removed.

Such oscillations, familiar from the staggered formulation, come about
since the underlying fermion field can create several different
species, and these species occur in different regions of the Brillouin
zone. It would be interesting to explore whether or not these
oscillations could be cancelled by constructing hadronic operators spread
over nearby neighbours \cite{Creutz:2010qm}.

It is important to remember that because the two species are of
opposite chirality, the naive $\gamma_5$ matrix is physically a flavour
non-singlet.  The naive on-site pseudoscalar field
$\overline{\psi}\gamma_5\psi$ can create only flavour non-singlet
pseudoscalar states.  To create the flavour-singlet pseudoscalar meson,
which gets its mass from the anomaly, one needs to combine fields on
nearby sites with appropriate phases.

We would like to stress that the breaking of hyper-cubic symmetry 
does not generate any sign problem for the Monte Carlo generation 
of configurations. The gauge action is real, and the eigenvalues 
of the Dirac operator come in complex conjugate pairs, so that 
the fermion determinant is always non-negative.

The purely gluonic counterterm for Bori\c{c}i-Creutz fermions 
introduces in the renormalized action operators of the kind 
$E\cdot B$, $E_1 E_2$, $B_2 B_3$ (and similar).
In a hyper-cubic invariant theory, instead, only the standard terms 
$E^2$ and $B^2$ are allowed.
Fixing the coefficient $c_P$ could then be done by measuring 
$\langle E\cdot B \rangle$, $\langle E_1 E_2 \rangle$, $\cdots$, 
and tuning $c_P$ in such a way that one (or more) of these 
expectation values is restored to its proper value pertinent to 
a hyper-cubic invariant theory, i.e. zero.
These effects could turn out to be rather small, given that only the 
fermionic part of the tree-level action breaks hyper-cubic symmetry.
It could also be that other derived quantities are more sensitive
to this coefficient, and more suitable for its nonperturbative determination.
In general one can look for Ward identities in which violations 
of the standard Lorentz invariant form, as functions of $c_P$, occur.

For Karsten-Wilczek fermions the purely gluonic counterterm introduces
an asymmetry between Wilson loops containing temporal links relative
to those involving spatial links only. One could then fix $d_P$ by computing 
a Wilson loop lying entirely in two spatial directions, and then equating 
its result to an ordinary Wilson loop which also has links in the time 
direction.

In the end only Monte Carlo simulations will reveal the actual amount 
of symmetry breaking. This could turn out to be large or small depending 
on the observable considered. 
One important such quantity is the mass splitting of the charged 
pions relative to the neutral pion.
Indeed, since there is only a $U(1) \otimes U(1)$ 
chiral symmetry, the $\pi^0$ is massless, as the unique Goldstone boson
(for $m_0\to 0$), but $\pi^+$ and $\pi^-$ are massive.

Furthermore, the magnitude of these symmetry-breaking effects could turn out
to be substantially different for Bori\c{c}i-Creutz compared to Karsten-Wilczek
fermions. Thus, one of these two actions could in this way be raised to
become the preferred one for numerical simulations.

\section{A unifying notation for the two fermion discretizations}
\label{sec:notation}

By introducing a particular notation, some similarities between 
the two realizations of minimally doubled fermions can be revealed.
This applies to the form of the action, operators and counterterms.
For this purpose one can introduce a 4-component object $\Lambda_\mu$, 
defined as
\begin{equation}
  \Lambda_{\mu} \equiv  \left\{\begin{array}{ll}
  \delta_{\mu 4} & \,\mbox{Karsten-Wilczek} \\[0.2cm]
  \frac{1}{2} & \,\mbox{Bori\c{c}i-Creutz}       
  \end{array}\right.,
  \qquad
  (\Lambda\cdot\gamma) \equiv \left\{\begin{array}{ll}
  \gamma^4 & \,\mbox{Karsten-Wilczek} \\[0.2cm]
  \Gamma & \,\mbox{Bori\c{c}i-Creutz}       
  \end{array}\right. .
\end{equation}
In both cases this object points from the zero of the action at the center 
of the Brillouin zone to the other zero (describing the second fermion, 
of opposite chirality).

At first we show that by means of this object one can cast both actions 
into similar (although non-equivalent) forms. Their free Dirac operators, 
as we have already seen in Section \ref{sec:simulations}, contain the same 
naive fermion piece but a different dimension-five operator. The latter 
can be rewritten in this new notation as
\begin{eqnarray}
  D_{KW}^{(5)}(k) &\equiv&
  \frac{2i}{a} \,\sum\limits_{\mu,\nu}\,
  \Lambda^{\nu} \gamma^{\nu} \,\sin^2 \frac{ap_{\mu}}{2}
    \, \Big(1-\delta_{\mu \nu }\Big) , \\
  D_{BC}^{(5)}(k) &\equiv& 
  -\frac{2i}{a} \,\sum\limits_{\mu,\nu}\,
  \Lambda^{\nu} \gamma^{\nu} \,\sin^2 \frac{ap_{\mu}}{2}
   \, \Big(1-2\delta_{\mu \nu}\Big) .
\end{eqnarray}
The factors $(1-\delta_{\mu \nu})$ and $(1-2\delta_{\mu \nu})$ cannot be 
transformed into each other, and this illustrates that the two actions 
are inequivalent and must be distinguished
(as we remarked in Section \ref{sec:actions}).

Although the quark propagator cannot be cast into a uniform expression
using this notation, this turns out to be possible for operators 
(e.g. local currents and counterterms), as well as some other results 
such as the expression for vacuum polarization. For example, 
the various counterterms that we have previously discussed can be easily
cast in a completely unified way for the two actions. If we rewrite the three 
counterterms making use of the object $\Lambda_\mu$,
the counterterms of dimension three appear as
\begin{equation}
i \overline{\psi}(x) (\Lambda\cdot\gamma) \psi(x) ,
\end{equation}
the fermionic ones of dimension four become
\begin{equation}
\overline{\psi}(x) (\Lambda\cdot\gamma)(\Lambda\cdot D) \psi(x) ,
\end{equation}
and the gluonic ones are
\begin{equation}
\sum_{\mu,\nu, \rho} \Lambda_{\mu} F_{\mu\rho} F_{\rho\nu} \Lambda_{\nu} .
\end{equation}
Here (and in the following) objects written in this unified notation 
may differ by simple numerical coefficients from the corresponding 
quantities which we have previously used in the conventional notation.

Let us now consider the results of the one-loop calculation that we
have presented in the previous Sections. One can rewrite 
the full self-energy (without counterterms) for both actions as
\begin{equation}
 \Sigma = i\slash{p} \,\Sigma_1 + m_0 \,\Sigma_2 
 + i\tilde{c}_1 \,(\Lambda\cdot\gamma)(\Lambda \cdot p)
 + \tilde{c}_2 \,\frac{i}{a}(\Lambda\cdot\gamma) ,
\end{equation}
with $\tilde{c}_i$ being given by either $c_i$ or $d_i$.
Also the fermionic bilinears can be expressed in a unified form.
Using the abbreviations $b=\frac{g_{0}^{2}C_{F}}{16\pi^{2}}$ and 
$L = \log a^{2}p^{2}$, the results for the one-loop vertex diagram
for the local scalar, vector and tensor bilinears are
\begin{eqnarray}
  C^{S} &=&
  b \,\left\{\begin{array}{ll}
  \Big(-4L+24.36875 + (1-\alpha)\big(L-5.792010\big) 
    \Big)~~\mbox{Karsten-Wilczek} \\
  \Big(-4L+29.48729 + (1-\alpha)\big(L-5.792010\big) 
    \Big)~~\mbox{Bori\c{c}i-Creutz}
  \end{array}\right. , \\
  C^{V}_{\mu} &=&
  b \,\left\{\begin{array}{ll}
  \gamma_{\mu}\Big(-L+10.44610+(1-\alpha)\big(L-4.792010\big)\Big) 
     -2.88914\cdot\Lambda_{\mu}(\Lambda\cdot\gamma)~~\mbox{Karsten-Wilczek} \\
  \gamma_{\mu}\Big(-L+9.54612+(1-\alpha)\big(L-4.792010\big)\Big) 
     -0.20074\cdot\Lambda_{\mu}(\Lambda\cdot\gamma)~~\mbox{Bori\c{c}i-Creutz}
  \end{array}\right. , \\
  C^{T}_{\mu\nu} &=&
  b \,\left\{\begin{array}{ll}
  \sigma_{\mu\nu}\Big(4.17551+(1-\alpha)\big(L-3.792010\big)
       \Big)~~\mbox{Karsten-Wilczek} \\
  \sigma_{\mu\nu}\Big(2.16548+(1-\alpha)\big(L-3.792010\big)
       \Big)~~\mbox{Bori\c{c}i-Creutz}
  \end{array}\right. .
\end{eqnarray}
For the conserved vector current, the sum of the standard proper diagrams 
(vertex, sails and operator tadpole) reads for the two actions
\begin{equation}
  b\,\left\{\begin{array}{ll}
\gamma_{\mu}\Big(-L-9.24089 + (1-\alpha)\big(L-4.792010\big) 
  +0.12554\cdot\Lambda_{\mu}(\Lambda\cdot\gamma)~~\mbox{Karsten-Wilczek} \\
\gamma_{\mu}\Big(-L-6.80663 + (1-\alpha)\big(L-4.792010\big)
 \Big)
  -3.05532\cdot\Lambda_{\mu}(\Lambda\cdot\gamma)~~\mbox{Bori\c{c}i-Creutz}
  \end{array}\right. .
\end{equation}

Perhaps one of the most striking examples of the convenience 
of this notation can be observed in the case of the vacuum polarization. 
The contribution of fermion loops to this quantity contains structures 
which break hyper-cubic symmetry. It can be written as
\begin{equation}
  \Pi^{(f)}_{\mu\nu}(p) = \Sigma_ {3}\ (p_\mu p_\nu-p^{2}\delta_{\mu\nu})
   +d_{g} \Big((\Lambda\cdot p)(\Lambda_{\mu}p_{\nu}+\Lambda_{\nu}p_{\mu})
     -(\Lambda_{\mu}\Lambda_{\nu}p^{2}+\delta_{\mu\nu}(\Lambda\cdot p)^{2})\Big) ,
\label{eq:vacpol}
\end{equation}
with the numerical results (as we have seen in Section \ref{sec:detgluon})
\begin{eqnarray}
  \Sigma_{3}(g_{0}^{2}) &=& \tilde{b}\,\left\{ \begin{array}{ll}
  -\frac{8}{3}L+19.99468~~\mbox{Karsten-Wilczek} \\
  -\frac{8}{3}L+23.6793~~\mbox{Bori\c{c}i-Creutz}
  \end{array}\right. , \\
  d_{g}(g_{0}^{2}) &=& \tilde{b}\,\left\{ \begin{array}{ll}
  -12.69766~~\mbox{Karsten-Wilczek} \\
  -3.6376~~\mbox{Bori\c{c}i-Creutz}       
  \end{array}\right.,
\end{eqnarray}
with $\tilde{b}= \frac{g_{0}^{2}C_2}{16\pi^{2}}$ (Wilson fermions have
$\Sigma_{3} = \tilde{b}(-\frac{4}{3}L+4.337002)$ and $d_{g}= 0$).
Thus, a single formula can describe the structures which arise
in the calculation of the vacuum polarization for both actions.

With this notation we have thus shown that operator structures and results 
for Bori\c{c}i-Creutz and Karsten-Wilczek fermions, although distinct, 
share many common traits.
As can be inspected in the above expressions, another remarkable feature 
appears to be that, after $\Lambda_\mu$ is introduced, the summed indices 
occur in pairs (like in the continuum), and also the free indices match 
exactly on both sides of equations. We do not know if this will always happen, 
also if one computes more complicated quantities.

Even without using the $\Lambda$ notation, we also discovered that the 
hyper-cubic-breaking terms of the vacuum polarization in eq.~(\ref{eq:vacpol}) 
can be put for both actions in the same algebraic form, namely
\begin{equation}
p^2 \{\gamma_\mu,\Gamma\} \{\gamma_\nu,\Gamma\}
+ \delta_{\mu\nu} \{\slash{p},\Gamma\}\{\slash{p},\Gamma\}
-\frac{1}{2}\,\{\slash{p},\Gamma\}
\Big( \{\gamma_\mu,\slash{p}\} \{\gamma_\nu,\Gamma\}
     +\{\gamma_\nu,\slash{p}\} \{\gamma_\mu,\Gamma\} \Big) ,
\end{equation}
where in the case of Karsten-Wilczek fermions $\Gamma$ must be
replaced by $\gamma_4/2$. This substitution is suggested by comparison
of the standard relation $\Gamma =
\frac{1}{4}\,\sum_{\mu}(\gamma_\mu+\gamma'_\mu)$ of Bori\c{c}i-Creutz
fermions with the formula $\gamma_4 =
\frac{1}{2}\,\sum_{\mu}(\gamma_\mu+\gamma'_\mu)$ for Karsten-Wilczek fermions, 
expressing the symmetries of the action 
(as can be seen from the expression of the propagator, when one expands 
it around the second zero).  
Whether there is any deeper significance to this structural ``equivalence'' 
of the hyper-cubic-breaking structures in the vacuum polarizations remains 
an open question.

\section{Conclusions}

Bori\c{c}i-Creutz and Karsten-Wilczek fermions are described by a fully 
consistent renormalized quantum field theory.
Three counterterms need to be added to the bare actions, and 
all their coefficients can be calculated either in perturbation theory
(as we have shown), or nonperturbatively from Monte Carlo simulations
(a task for the future, for which we have suggested some strategies).
After these subtractions are consistently taken into account, the
power divergence in the self-energy is eliminated, and 
no other power divergences occur for all quantities that we calculated.

We have argued that under reasonable assumptions and following the 
nonperturbative determination of these counterterms, no special features 
of these two realizations of minimally doubled fermions should hinder their 
successful Monte Carlo simulation.

Conserved vector and axial currents can be derived, and they have simple
expressions which involve only nearest-neighbours sites. 
We have then here one of the very few cases where one can define 
a simple conserved axial current (also ultralocal).

Finally, we would like to observe that this work is also an example 
of the usefulness of perturbation theory in helping to unfold theoretical 
aspects of (new) lattice formulations.

\acknowledgments

This work was supported by Deutsche Forschungsgemeinschaft
(SFB443), the GSI Helmholtz-Zentrum f\"ur Schwerionenforschung, and
the Research Centre ``Elementary Forces and Mathematical Foundations''
(EMG) funded by the State of Rhineland-Palatinate. MC was supported
by contract number DE-AC02-98CH10886 with the U.S.~Department of
Energy.  Accordingly, the U.S. Government retains a non-exclusive,
royalty-free license to publish or reproduce the published form of
this contribution, or allow others to do so, for U.S.~Government
purposes.  MC is particularly grateful to the Alexander von Humboldt
Foundation for support for multiple visits to the University of Mainz.

\end{document}